\documentclass[twocolumn,secnumarabic,amssymb, nobibnotes, aps,
prd]{revtex4} \usepackage{graphicx}

\begin{document}

\title{Differences between application of some basic principles of quantum
mechanics on atomic and mesoscopic levels}

\author{Alexey Nikulov}


\affiliation{Institute of Microelectronics Technology and High Purity
Materials, Russian Academy of Sciences, 142432 Chernogolovka, Moscow
District, RUSSIA. nikulov@ipmt-hpm.ac.ru}


\begin{abstract} Formalism of the quantum mechanics developed for
microscopic (atomic) level comes into collision with some logical
difficulties on mesoscopic level. Some fundamental differences between
application of its basic principles on microscopic and  mesoscopic levels
are accentuated.
\end{abstract}

\maketitle

\narrowtext

\section{Introduction}
Richard Feynman remarked: {\it "I think I can safely say that nobody today
understands quantum physics"}. This remark may seem queer for people who
studied and use quantum physics but some experts understand that {\it in
contrast to the theories of relativity, quantum mechanics is not yet based
on a generally accepted conceptual foundation} \cite{Nikulov01}. Not only
the collision of principles of quantum mechanics with macroscopic realism
\cite{Nikulov02,Nikulov03} and the Einstein-Podolsky-Rosen paradox
\cite{Nikulov04,Nikulov05} are indicative of our incomprehension of quantum
physics. There are some quantum effects observed, first of all, on the
mesoscopic level, strangeness of which is disregarded by most scientists
who do not understand that {\it nobody today understands quantum physics}.

The experimental results corroborate for the present all principles of
quantum physics \cite{Nikulov06}, even in defiance of common sense
\cite{Nikulov07,Nikulov08}. But the essence of these principles is not
clear and is discussed now actively \cite{Nikulov09,Nikulov10,Nikulov11}.
The collision between quantum mechanics and macroscopic realism
\cite{Nikulov02,Nikulov03} should be expected on the mesoscopic level.
Therefore the consideration of differences between application of basic
principles of quantum mechanics on atomic and mesoscopic levels is most
urgent.

\section {Quantum mechanics versus macroscopic realism}

One of the three "axises" along which, according to A.J. Leggett
\cite{Nikulov12}, it is not unreasonable to seek evidence of a breakdown of
the quantum mechanics scheme of the physical world is the collision of it
with our immediate experience of the "everyday" world. The obvious
contradiction between the quantum mechanics and macroscopic realism was
laid stress by Erwin Schrodinger already seventy years ago \cite{Nikulov13}
but only in the last years this problem is not only merely philosophical
but it can be tested in experiment \cite{Nikulov03} first of all on the
mesoscopic level, i.e. between the microscopic (atomic) world and the
Schrodinger cat. The formalism of the quantum mechanics, its Copenhagen
interpretation, was developed first of all for the microscopic (atomic)
level and it comes into collision with some logical difficulties on the
mesoscopic level.

According to the formalism of the quantum mechanics a quantum system can be
in a superposition of states but this superposition can not be observed
because of its reduction to single state at measuring. The principle of the
impossibility of noninvasive measurement seems admissible on the
microscopic level when measuring device can not be smaller than measured
object. But we can not assume that the Schrodinger cat can die or revive
because of our look. The contradiction between quantum mechanics and the
possibility of noninvasive measurability \cite{Nikulov02,Nikulov14} may can
emerge on the mesoscopic level.

\section{Quantization of the momentum circulation}

Other difficulty can be connected with the quantization of momentum
circulation. According to the classical physics the momentum $p = mv + qA$
of a particle with a charge $q$ should maintain a constant value in absence
of any force whereas the quantum number $n$ in the relation for the
momentum circulation $$\oint_{l}dl p =  \oint_{l}dl (mv + qA) =
m\oint_{l}dlv + q\Phi = n2\pi \hbar \eqno{(1)} $$ can change without any
evident force. There is not problem on the microscopic realm, where
electrons do not change their state of motion in the absence of an
electromagnetic force but the problem is on the mesoscopic level
\cite{Nikulov15}. The 'mysterious' change of state of electron motion
without forces acting on the electrons can be both in superconductor
\cite{Nikulov15} and other (semiconductor and normal metal) mesoscopic
structures with the quantization (1) of momentum circulation.

The quantization (1) takes place $\oint_{l}dl p = n2\pi \hbar$ when the
wave function of a particle is closed in a two-connected mesoscopic loop
and $m\oint_{l}dlv = n2\pi \hbar - q\Phi = 2\pi \hbar(n - \Phi/\Phi_{0})
\neq 0$, i.e. the state with zero velocity $v = 0$ is forbidden, when the
magnetic flux $\Phi$ inside the loop is not divisible by the flux quantum
$\Phi_{0} = 2\pi \hbar/q$. On the other hand the velocity can be zero $v =
0$ in the state with unclosed wave function when the quantization (1) is
not valid.  In this case the circular velocity of the particle $v$ should
change, i.e. the particle should accelerate, from $v = 0$ to $v =
\oint_{l}dlv/l = 2\pi \hbar(n - \Phi/\Phi_{0})/l$ and the momentum
circulation should change from $q\Phi$ to $n2\pi \hbar$ at the closing of
the wave function without any evident force.

There is important to accentuate a fundamental difference between atomic
and mesoscopic levels. A switching between states with different
connectivity of wave function can not be realized on atomic level whereas
it can be enough easy made on mesoscopic level. For example it can be
realized by switching of a segment $l_{s}$ of a loop $l$ between
superconducting, i.e. with a density of superconducting pairs $n_{s} > 0$,
and normal states with $n_{s} = 0$, whereas other segment $l_{scs} = l -
l_{s}$ remaining all time in superconducting state with $n_{s} > 0$
\cite{JLTP98,PRB01}. The quantization (1) should be along any closed path
$l$ of the loop circumference when $n_{s} > 0$ along whole loop and the
quantization (1) is not valid along $l$ when $n_{s} = 0$ in the $l_{s}$
segment. The velocity of superconducting pairs $$\oint_{l}dlv_{s} =
\frac{2\pi \hbar}{m}(n - \frac{\Phi}{\Phi_{0}}) \eqno{(2)}$$ and a density
of the persistent current $j_{p} = 2en_{s}v_{s} \neq 0$ should be nonzero
along $l$ in the closed superconducting state at $\Phi \neq n\Phi_{0}$
because of the quantization (1), whereas equilibrium velocity $v_{s} = 0$
and current $j_{p} = 0$ in the $l_{scs}$ segment when the $l_{s}$ segment
is in the normal state with a non-zero resistance $R_{ls} > 0$. Thus,
superconducting pairs in the $l_{scs}$ segment should accelerate without
any force, in contradiction with the law of momentum conservation, at the
switching of the $l_{s}$ segment from the normal $n_{s} = 0$ to
superconducting $n_{s} > 0$ state. This change can be fixed  experimentally
by way of an observation of the appearance of  the persistent current at
closing of superconducting state.

The term "persistent current" was at first used for the current in
superconducting state \cite{SupPC61,SupPC63,SupPC65}, i.e. at $T < T_{c}$.
Under equilibrium conditions at $T < T_{c}$ the quantization (1) is valid
during all time since coherence of wave function of superconducting pairs
exists until the superconducting state exists. Above superconducting
transition $T > T_{c}$ superconducting pairs exist because of thermal
fluctuations \cite{Skocpo75} and coherence of their wave function along
whole loop $l$ appears only at times. It is enough in order the persistent
current, i.e. a direct circular current observed under equilibrium
conditions, exists not only at $T < T_{c}$ but also in non-superconducting
state at $T > T_{c}$ \cite{Kulik1}, when the resistance along $l$ is not
zero $R_{l} > 0$. First experimental evidence of the persistent current at
$R_{l} > 0$ in the fluctuation region $T \geq T_{c}$ is the Little-Parks
oscillations of the resistance of cylinder \cite{LitPar62} or loop
\cite{LitPar92} in magnetic field $R_{l}(\Phi/\Phi_{0})$.

The observation of the circular persistent current $I_{p}$ at a constant
magnetic field $d\Phi/dt = 0$ in a loop with a non-zero resistance $R_{l} >
0$ contradicts to the habitual knowledge according to which such current
should disappear without the Faraday's voltage $\oint_{l} dl E_{F} =
-d\Phi/dt = 0$ because of dissipation, $R_{l}I_{p} \neq 0$, during the time
of current relaxation $\tau_{RL} = L_{l}/R_{l}$. According to the
explanation \cite{PRB01} the persistent current does not disappear at
$R_{l} > 0$ since the velocity decrease because of the dissipation force is
compensated by the velocity change because of the quantization (1) at
closing of superconducting state at reiterate switching of the loop by
thermal fluctuations between superconducting states with different
connectivity. The explanation \cite{PRB01} of the observation of  the
persistent power $R_{l}I_{p}^{2} \neq  0$ as a fluctuation phenomenon is
natural since $I_{p} \neq 0$ at $R_{l} > 0$ is observed only in the
fluctuation region near $T_{c}$, where the loop is switched by fluctuations
between superconducting states with different connectivity. According to
this explanation \cite{PRB01} the observation of the persistent current
$I_{p} \neq 0$ at $R_{l} > 0$ in the fluctuation region of superconducting
loop is experimental evidence of violation of the law of conservation of
momentum circulation. Already the observation of the direct circular
current $I_{p}$ at $d\Phi/dt = 0$ and $R_{l} > 0$ is challenge to this law
since it is observed at $R_{l} > 0$, as well as a conventional circular
current, but without the circular Faraday's force $2eE_{F}$, $\oint_{l} dl
2eE_{F} = -2ed\Phi/dt = 0$.

The wave function not only superconducting pairs in the fluctuation region
at $T > T_{c}$ but also of electrons in mesoscopic semiconductor and normal
metal loops can become closed at times. I.O.Kulik predicted first the
persistent current in normal metal mesoscopic structure \cite{Kulik2} just
after the consideration of this quantum phenomenon at $T > T_{c}$ in
superconductor \cite{Kulik1}. It is much more difficult to observed the
persistent current of electron than superconducting pairs. Nevertheless the
advancement of cryogenic and microfabrication technologies had allowed to
make attempts to observe the persistent current in semiconductor
\cite{Semic93,Semic95,Semic01} and normal metal
\cite{Normal90,Normal91,Normal01} nanostructures. First it was made only in
1990, i.e.  in 20 years after the prediction\cite{Kulik2}. It may be
therefore most authors refer to \cite{Buttiker} as the first prediction of
the persistent current in non-superconducting structures. The persistent
current in non-superconducting loops also contradicts to the habitual
knowledge since the resistance of these loop is not zero.

An additional, more obvious, experimental evidence of  violation of the law
of momentum conservation is the observation \cite{Dub02,Nikulov16} of the
quantum oscillations of the dc voltage $V_{dc}(\Phi/\Phi_{0})$ on segments
of asymmetric superconducting loops predicted in \cite{JLTP98,PRB01}. The
potential difference $R_{ls}I_{p}$ should appear on the segment $l_{s}$
just after its switching in the normal state with $R_{ls} > 0$ if the
persistent current in the loop $l$ was non-zero $I_{p} \neq 0$ before the
switching. This potential difference $V(t) = R_{ls}I(t) = R_{ls}I_{p}
\exp(-t/\tau_{RL})$, as well as the circular current $I(t) = I_{p}
\exp(-t/\tau_{RL})$, are extinguished during a finite time of current
relaxation $\tau_{RL} = L_{l}/R_{ls}$ because of a finite value of the loop
inductance $L_{l}$. The time average of the $V(t)$ voltage  during the time
$t_{n}$ of a staying of the $l_{s}$ in the normal state
$\overline{V}^{t_{n}} = t_{n}^{-1}\int_{0}^{t_{n}}V(t) = R_{ls}I_{p}
t_{n}^{-1}\int_{0}^{t_{n}} \exp(-t/\tau_{RL})$ equals $\overline{V}^{t_{n}}
\approx R_{ls}I_{p}$ at $t_{n} \ll \tau_{RL}$ and $\overline{V}^{t_{n}}
\approx L_{l}I_{p}/t_{n}$ at $t_{n} \gg \tau_{RL}$. The dc component of the
voltage measured during a long time $\Theta$, $V_{dc} =
\Theta^{-1}\int_{\Theta }dtV(t) =
N_{sw}^{-1}\sum_{Nsw}R_{ls}I_{p}\omega_{sw} \int_{0}^{t_{n}}
\exp(-t/\tau_{RL})$ equals $V_{dc} \approx
\overline{R_{ls}I_{p}t_{n}}\omega_{sw}$ at $t_{n} \ll \tau_{RL}$ and
$V_{dc} \approx L_{l}\omega_{sw}\overline{I_{p}}$ at $t_{n} \gg \tau_{RL}$
in the case of reiterate switching of the $l_{s}$ segment between
superconducting and normal states with a frequency $\omega_{sw} =
N_{sw}/\Theta$.

The switching of the $l_{s}$ with the frequency $\omega_{sw}$ means that
during the long time $\Theta$ the loop $l$ is $N_{sw}$ times in the closed
superconducting state and $N_{sw}$ times in the unclosed superconducting
state. The density of the persistent current $j_{p} = 2en_{s}v_{s}$ in each
(from $N_{sw}$) closed superconducting state is determined by the $n_{s}$
value and the quantization of the velocity (2). The density  $j_{p}$ is
uniform across the narrow section $s \ll \lambda_{L}^{2}$ of the loops
measured in \cite{JLTP98,PRB01}. Where $\lambda_{L}$ is the London
penetration depth. The persistent current in the closed superconducting
state of the loop equals $I_{p} = sj_{p} = s2en_{s}v_{s} = (2e\pi \hbar
/lm<(sn_{s})^{-1}>) (n - \Phi/\Phi_{0})$ because of the quantization (2)
and since its value should be uniform along $l$ in the stationary state:
$\oint_{l}dl v_{s} = (I_{p}/2e)\oint_{l}dl (sn_{s})^{-1} =
(I_{p}/2e)l<(sn_{s})^{-1}>$. The quantum number $n$ can be any integer
number in the closed superconducting state but with overwhelming
probability $P_{n} \propto \exp(-E_{n}/k_{B}T)$ the loop switches in the
permitted state with lowest energy $E_{n}$ since the energy difference
$E_{n+1} - E_{n}$ between adjacent permitted states is much higher than the
thermal energy $k_{B}T$ \cite{PRB01}. Therefore the average value
$\overline{n} = N_{sw}^{-1}\sum_{Nsw}n = \sum_{n}nP_{n}$ is close to the
integer number corresponding to the lowest $v_{s}^{2} \propto (n -
\Phi/\Phi_{0})^{2}$ value and $\overline{I_{p}} =
N_{sw}^{-1}\sum_{Nsw}I_{p}$ is not zero at $\Phi \neq n\Phi_{0}$ and $\Phi
\neq (n+0.5)\Phi_{0}$. $\overline{I_{p}} = 0$ at  $\Phi \neq
(n+0.5)\Phi_{0}$ since two permitted states, $n - \Phi/\Phi_{0} = 1/2$ and
$n - \Phi/\Phi_{0} = -1/2$, with opposite direction of the persistent
current $I_{p} \propto n - \Phi/\Phi_{0}$ have the same energy $(n -
\Phi/\Phi_{0})^{2} = (1/2)^{2} =  (-1/2)^{2}$ and therefore $\overline{n} -
\Phi/\Phi_{0} = 1/2 + (-1/2) = 0$ at $\Phi \neq (n+0.5)\Phi_{0}$.

Thus, the dc voltage $V_{dc} \propto \overline{I_{p}} \propto \overline{n}
- \Phi/\Phi_{0}$, sign and value of which are periodical function of the
magnetic flux $V_{dc}(\Phi/\Phi_{0})$ should be observed on the $l_{s}$
segment at its reiterate switching between superconducting and normal
states. Just such quantum oscillations of the dc voltage
$V_{dc}(\Phi/\Phi_{0})$ were observed in \cite{Dub02,Nikulov16}. There is
important that the dc potential difference $V_{dc}(\Phi/\Phi_{0})$ is
observed both on the switched segment $l_{s}$ and other one $l_{scs} = l -
l_{s}$ remaining all time in superconducting state. The latter is possible
since the acceleration of pair in the electric field $\overline{dp/dt} =
2e\overline{E_{p}} = 2eV_{dc}/l_{scs}$ is equilibrated by the momentum
change, i.e. by the acceleration in opposite direction \cite{JLTP98,PRB01},
because of the quantization (1). The momentum circulation $\oint_{l}dl p$
of superconducting pair with the charge $q = 2e$ changes from $2e\Phi$ to
$n2\pi \hbar$ at each closing of the wave function. The average value of
this change $N_{sw}^{-1}\sum_{Nsw}(2\pi \hbar n - 2e\Phi) = 2\pi \hbar
(\overline{n} - \Phi/\Phi_{0})$ depends periodically on magnetic flux as
well as the dc voltage $V_{dc}(\Phi/\Phi_{0}) \propto (\overline{n} -
\Phi/\Phi_{0})$ observed in \cite{Dub02,Nikulov16}. The observation of the
dc voltage on the $l_{scs}$ segment remaining all time in superconducting
state contradicts to the law of momentum conservation.  The quantum
oscillation of the dc voltage $V_{dc}(\Phi/\Phi_{0})$ may be expected also
in semiconductor and normal metal asymmetric mesoscopic loops.

\section{Intrinsic breach of symmetry}

The law of momentum conservation is connected with symmetry of space and
the violation of this law at the closing of the wave function can be
connected with the intrinsic breach of symmetry. The experimental evidence
of the intrinsic breach of symmetry is even more obvious than violation to
the law of momentum conservation. It is observed in  \cite{Dub02,Nikulov16}
that the potential electric field $E_{p}(\Phi/\Phi_{0}) = -\bigtriangledown
V(\Phi/\Phi_{0})$ has right or left direction which changes periodically
with the value $\Phi/\Phi_{0}$ of the magnetic flux. For example, if the
$E_{p}$ direction is right at $\Phi/\Phi_{0} = 1/4$ then it is left at
$\Phi/\Phi_{0} = 3/4$ \cite{Dub02,Nikulov16}. It is very strange that
direction of a vector changes with a scalar value. We should ask: "Why can
the dc electric field $E_{p}$ have right direction at $\Phi/\Phi_{0} = 1/4$
and left one at $\Phi/\Phi_{0} = 3/4$?" There can be only answer: "Because
the loop is asymmetric, for example the lower half is more narrow than the
upper one, see Fig.4 in \cite{Nikulov16}, and  the circular persistent
current has contra-clockwise direction at $\Phi/\Phi_{0} = 1/4$ and
clockwise one at $\Phi/\Phi_{0} = 3/4$".

It seems self-evident that any direct current has a direction. Nobody
doubts that a conventional direct circular current $I =
R_{l}^{-1}(-d\Phi/dt)$ (it is in the stationary regime at $t \gg
L_{l}/R_{l}$) induced in a loop with a resistance $R_{l}$ by the Faraday's
voltage $\oint_{l} dl E = -d\Phi/dt$ has clockwise or contra-clockwise
direction and this direction determines right or left direction of the
potential electric field $E_{p} = -\bigtriangledown V$ observed on a loop
segment $l_{s}$ the resistivity $R_{ls}/l_{s}$ of which differs from the
one $R_{l}/l$ along whole loop $l$, when $V = (R_{ls}/l_{s} -
R_{l}/l)l_{s}I$. But it is no so obvious for the persistent current
existing because of the Bohr's quantization, as well as stable electron
orbit in atom. There is important to accentuate the fundamental difference
of the persistent current, as one of the mesoscopic quantum phenomena, from
the conventional current, on the one hand, and from electron orbit in atom
(1), on the other hand.

The direction of a conventional  circular current is determined by the
circular Faraday electric field $\oint_{l} dl E = -d\Phi/dt$. But the
persistent current is observed at a constant magnetic flux $\Phi$ and,
according to the experimental evidence \cite{Dub02,Nikulov16}, its
direction changes with a scalar value $\Phi/\Phi_{0}$ without any external
vector factor, i.e. the $I_{p}$ can have different directions at the same
direction of the magnetic flux $\Phi$ when the  $\Phi$ values are
different. The observation \cite{Dub02,Nikulov16} of a direction of the
persistent current is experimental evidence of intrinsic breach of
clockwise - counter-clockwise symmetry, since, in contrast to the
conventional circular current, the $I_{p}$ direction is not determined by
an external vector. The periodical dependence  $I_{p}(\Phi/\Phi_{0})
\propto V_{dc}(\Phi/\Phi_{0})$ of the direction of the persistent current
with the period $\Phi_{0} = 2\pi \hbar/q$ is indubitable evidence that this
intrinsic breach of symmetry is consequence of the Bohr's quantization (1).

Bohr postulated the quantization (1), $\oint_{l}dl p = \oint_{l}dl mv =
n2\pi \hbar$ at $\Phi = 0$, in order to explain the stability of electron
orbit in atom. There was a  logical difficulty in this model until electron
considered as a particle having a velocity $v$ since it was impossible to
answer on the question: "What direction has the velocity of electron on
stable atomic orbit?" The uncertainty relation $\Delta p\Delta l \geq \hbar
$ and the wave quantum mechanics have overcome this difficulty. Electron
can not has a certain coordinate on stable atomic orbit with a certain
momentum according to the uncertainty relation and therefore it can not
have a velocity. It is a wave but not a particle in the case of the Bohr's
quantization on atomic orbit. Therefore the Bohr's quantization  does not
break a symmetry on the atomic level. But we see that the breach of
symmetry because of the Bohr's quantization is observed
\cite{Dub02,Nikulov16} on the mesoscopic level.

This intrinsic breach of symmetry is observed since the canonical momentum
$p = mv + qA$ includes not only velocity $v$ but also a magnetic vector
potential $A$ and therefore sign and value of a circular velocity on the
lowest permitted state (2) depend periodidically on the $\Phi/\Phi_{0}$. It
may be considered as the cause of the periodical changes of equilibrium
magnetization $M(\Phi/\Phi_{0})$ of both superconductor and
non-superconductor, semiconductor \cite{Semic01} and normal metal
\cite{Normal01}, mesoscopic loops. It is very difficult to investigate
experimentally a possibility of like oscillations on atomic level since the
Bohr's radius, a typical atomic size $r_{B} \approx  0.053 \ nm$, is much
smaller than a radius $r_{B} = 500 \ nm$ of the mesoscopic loops. The very
high magnetic field $B > \Phi_{0}/\pi r_{B}^{2} \approx 3 \ 10^{5} \ T$ is
needed in order to observe the $M(\Phi/\Phi_{0})$ oscillations on atomic
level. It is important to note that the $M(\Phi/\Phi_{0})$ oscillations is
challenge to the law of momentum conservation since this periodical change
is evidence of change of the quantum number $n = \oint_{l}dl p/2\pi \hbar $
determining the value of momentum circulation (1). One may assume that this
change can be only at a breach of the coherence of wave function along $l$.

The intrinsic breach of symmetry on the mesoscopic level because of the
Bohr's quantization is challenge to some basic principle of statistical
mechanics and thermodynamics \cite{FQMT04} since it violates the postulate
of absolute randomness of any equilibrium motion \cite{QI2002}.

The work was financially supported by the Presidium of Russian Academy of
Sciences in the Program "Low- Dimensional Quantum Structures", by Russian
Foundation of Basic Research (Grant 04-02-17068) and by ITCS department of
Russian Academy of Sciences in the Program "Technology Basis of New
Computing Methods".

\end{document}